\documentclass[prl,reprint,twocolumn,superscriptaddress,noshowpacs,notitlepage,longbibliography]{revtex4-1}%
\usepackage{graphicx,bm,times}
\usepackage{amsmath}
\usepackage{amsfonts}
\usepackage{amssymb}
\usepackage{color}
\usepackage{hyperref}
\hypersetup{
	colorlinks = true,
	allcolors = {blue}
}

\begin{document}
	
	\title{Scaling of the superconducting gap with orbital character in FeSe}
	
	\author{Luke C. Rhodes}
	\affiliation{Diamond Light Source, Harwell Campus, Didcot, OX11 0DE, United Kingdom}
	\affiliation{Department of Physics, Royal Holloway, University of London, Egham, Surrey, TW20 0EX, United Kingdom}
	
	\author{Matthew D. Watson}
	\email[corresponding author:]{mdw5@st-andrews.ac.uk}
	\affiliation{School of Physics and Astronomy, University of St. Andrews, St. Andrews KY16 9SS, United Kingdom}
	\affiliation{Diamond Light Source, Harwell Campus, Didcot, OX11 0DE, United Kingdom}	

	\author{Amir A. Haghighirad}
	\affiliation{Clarendon Laboratory, Department of Physics,
		University of Oxford, Parks Road, Oxford OX1 3PU, United Kingdom}
	\affiliation{Institute for Solid State Physics, Karlsruhe Institute of Technology, 76021 Karlsruhe, Germany}
	
	\author{Daniil V. Evtushinsky}
	\affiliation{Institute of Physics, Ecole Polytechnique Federale Lausanne,CH-1015 Lausanne, Switzerland}

	\author{Matthias Eschrig}
	\affiliation{Department of Physics, Royal Holloway, University of London, Egham, Surrey, TW20 0EX, United Kingdom}
			
	\author{Timur K. Kim}
	\email[corresponding author:]{timur.kim@diamond.ac.uk}
	\affiliation{Diamond Light Source, Harwell Campus, Didcot, OX11 0DE, United Kingdom}

	\begin{abstract}
		 We use high-resolution angle-resolved photoemission spectroscopy to map the three-dimensional momentum dependence of the superconducting gap in FeSe. We find that on both the hole and electron Fermi surfaces, the magnitude of the gap follows the distribution of $d_{yz}$ orbital weight. Furthermore we theoretically determine the momentum dependence of the superconducting gap by solving the linearized gap equation using a tight binding model which quantitatively describes both the experimental band dispersions and orbital characters. By considering a Fermi surface only including one electron pocket, as observed spectroscopically, we obtain excellent agreement with the experimental gap structure.  Our finding of a scaling between the superconducting gap and the $d_{yz}$ orbital weight supports the interpretation of superconductivity mediated by spin-fluctuations in FeSe. 
	\end{abstract}
	\date{\today}
	\maketitle
	

Over the last ten years of extensive studies on the iron-based superconductors, many experimental works have provided support for the spin-fluctuation pairing hypothesis, including the observation of a spin-resonance peak in the superconducting (SC) state \cite{Inosov2009,Ma2017}, the presence of nodes in the SC gap in some systems \cite{Hashimoto2012}, evidence for a sign-change of the SC gap from quasi-particle interference \cite{HAEM2015}, as well as the more general observation of SC in close proximity to antiferromagnetism in the phase diagrams \cite{Hirschfeld2011}. 
A distinctive feature of spin-fluctuation mediated SC is that the pairing interactions are sensitive to the orbital character of the bands \cite{Graser2009,Oles1983,Takimoto2004}. Fundamentally, this sensitivity arises since the relevant interaction is the strong, local, and instantaneous Coulomb repulsion, which will generally be larger for the pairing of electrons in the same orbital \cite{Chubukov2012}.  Thus evidence for pairing by spin fluctuations in the Fe-based SC would be the observation of a direct relationship between the orbital character and the SC gaps around the Fermi surface. In most Fe-based SC, the complexities of the multiband, multiorbital electronic structures makes the experimental verification of this crucial link challenging. However here we focus on FeSe, which exhibits both nematic order and SC \cite{Bohmer2017}, and is known to have a relatively simple but highly anisotropic Fermi surface \cite{Watson2015,Watson2016,Fedorov2016,Kreisel2017a}, allowing detailed testing of the relationship between orbital character and SC \cite{Kang2018_arxiv,Sprau2017,Liu2018_arxiv}. 

\begin{figure*}
	\centering
	\includegraphics[width=\linewidth]{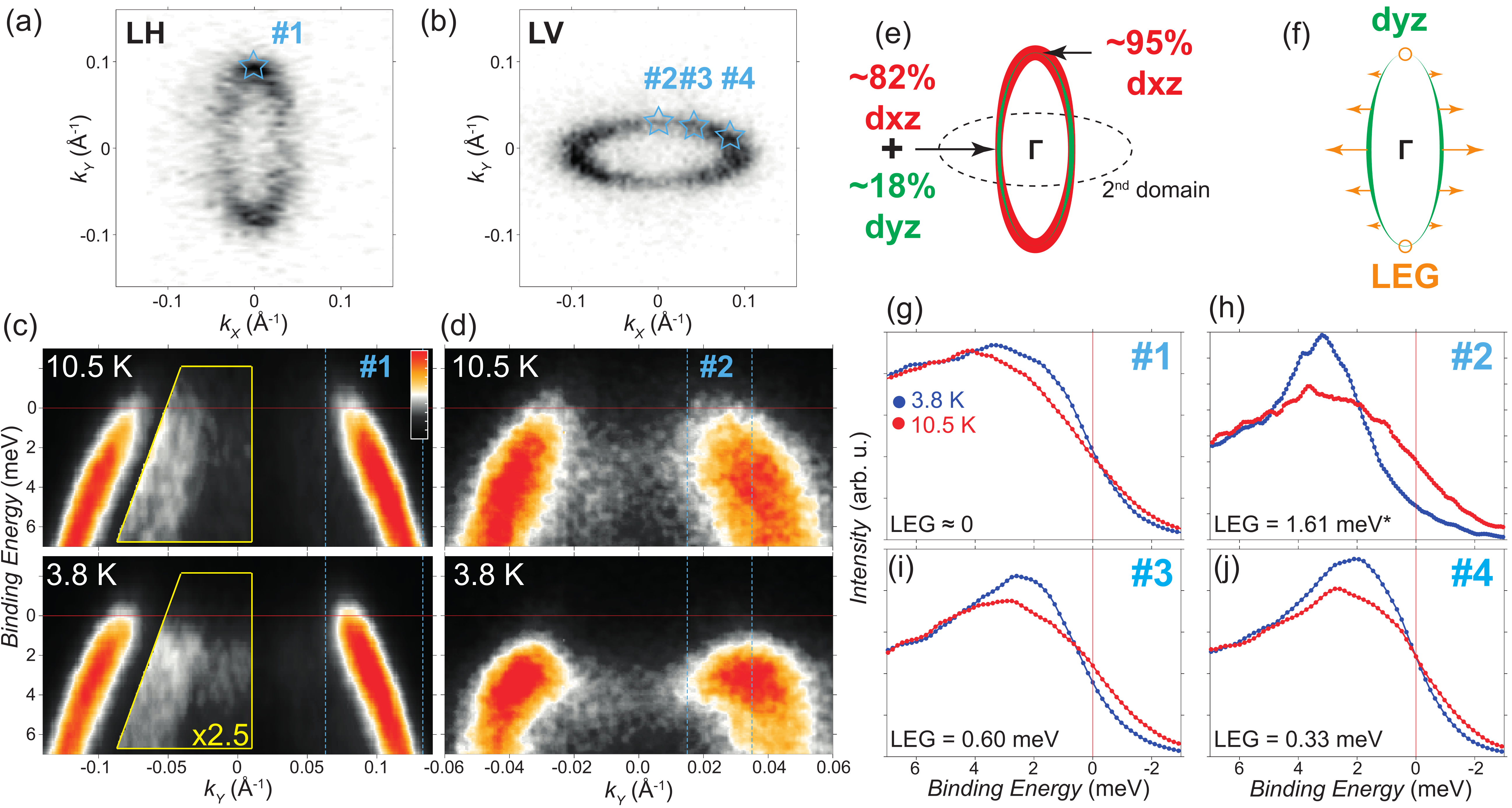}
	\caption{a,b) Fermi surface maps of twinned FeSe samples around the $\Gamma$ point (37 eV), in both linear polarizations. c) High-symmetry dispersions above and below $T_c$. d) High symmetry dispersion in LV polarization, highlighting the shorter axis of the ellipse. e) Schematic of the distribution of orbital weights. f) Schematic correlation between the LEG and the $d_{yz}$ orbital character. g,h) EDCs integrated in small regions (cyan dashed lines in (c,d)) around $k_F$.  i,j) EDCs at positions shown in (b), off the high-symmetry axes. *Note that (d) and (h) are obtained with a higher resolution than other plots.}
	\label{fig:fig1}
\end{figure*}

In this letter we present high resolution angle resolved photoemission spectroscopy (ARPES) measurements of the momentum-dependence of the SC gap. We show that the magnitude of the gap on both the hole and electron pockets follows the symmetry of the $d_{yz}$ orbital weight, while Fermi surface segments with predominantly $d_{xz}$ or $d_{xy}$ orbital character do not show observable gaps. We then present the theoretical solution to the linearized SC gap equation for a tight binding model that quantitatively reproduces not only the band dispersions observed in ARPES in the nematic phase but also the orbital characters. In addition we consider a scenario where the second electron pocket, which is expected to exist but is not observed spectroscopically \cite{Watson2017d,Sprau2017}, is incoherent and does not contribute to the SC pairing. In this case the anisotropy of the superconducting gap directly scales with the $d_{yz}$ orbital weight on both pockets, in excellent agreement with our experimental results.
Our direct observation of a simple relationship between $d_{yz}$ orbital weight and the SC spectral gap can be considered as a strong indication of the efficacy of pairing by spin fluctuations in an Fe-based SC, while the result also highlights the critical impact of an incoherent electron pocket in the nematic phase of FeSe.

Single crystals were grown by chemical vapor transport \cite{Watson2015}. ARPES measurements were performed at the I05 beamline at Diamond Light Source \cite{I05beamlinepaper}. The base temperature reached was 3.7~K at the sample position. In the best case, we achieve a total instrumental resolution of approximately 2~meV at 37~eV photon energy, with a more typical resolution of 3~meV.

In Fig.~\ref{fig:fig1}(a,b), we present Fermi surface maps obtained at 37~eV, the appropriate $k_z$ for the $\Gamma$ point where the hole pocket is smallest. The sample is not strained and is twinned, so a superposition of signals from the elliptical hole pocket from two nematic domains is expected. Strikingly, however, one can select which ellipse is primarily observed by switching the polarization. The matrix elements at the zone center are determined by the parity of the orbitals \cite{Brouet2012}; linear horizontal polarization (LH) couples strongly to $d_{xz}$ orbitals, whereas linear vertical polarization (LV) highlights the $d_{yz}$. The switching behavior implies that the hole pocket must be dominated by $d_{xz}$ character \footnote{In the nematic phase, we specify $x$ and $y$ to be along the $a$ and $b$ ($a>b$) axes respectively}. For the domain with $a$ oriented horizontally (and where the ellipse is elongated along $b$ \cite{Watson2017d}), this $d_{xz}$ character couples strongly to LH and therefore the vertical ellipse is observed. For the second domain, the reference frame of the orbitals is rotated, and therefore it is mainly the horizontal ellipse which is observed in LV polarization. The dominance of $d_{xz}$ orbital character and the pronounced ellipticity of the pocket are both direct consequences of nematic order \cite{Watson2016}, since in the tetragonal phase the pocket is circular with fourfold-symmetric $d_{xz}$ and $d_{yz}$ contributions. The dominance of $d_{xz}$ orbital character is particularly pronounced at the $\Gamma$ point compared with the Z point since here the pocket is smallest. Although the $d_{xz}$ is dominant, the switching effect is not perfect, for instance the inner band can also be observed weakly in the LH polarization high-symmetry cuts in Fig.~\ref{fig:fig1}(c). Quantitative analysis of the polarization switching effect (Supplemental Information, (SI)) suggests that the pocket is $\sim$95$\%$ $d_{xz}$ on the major axis and 82$\%$-18$\%$ $d_{xz}$-$d_{yz}$ on the minor axis, summarized in Fig.~\ref{fig:fig1}(e). 

\begin{figure*}[th]
	\centering
	\includegraphics[width=\linewidth]{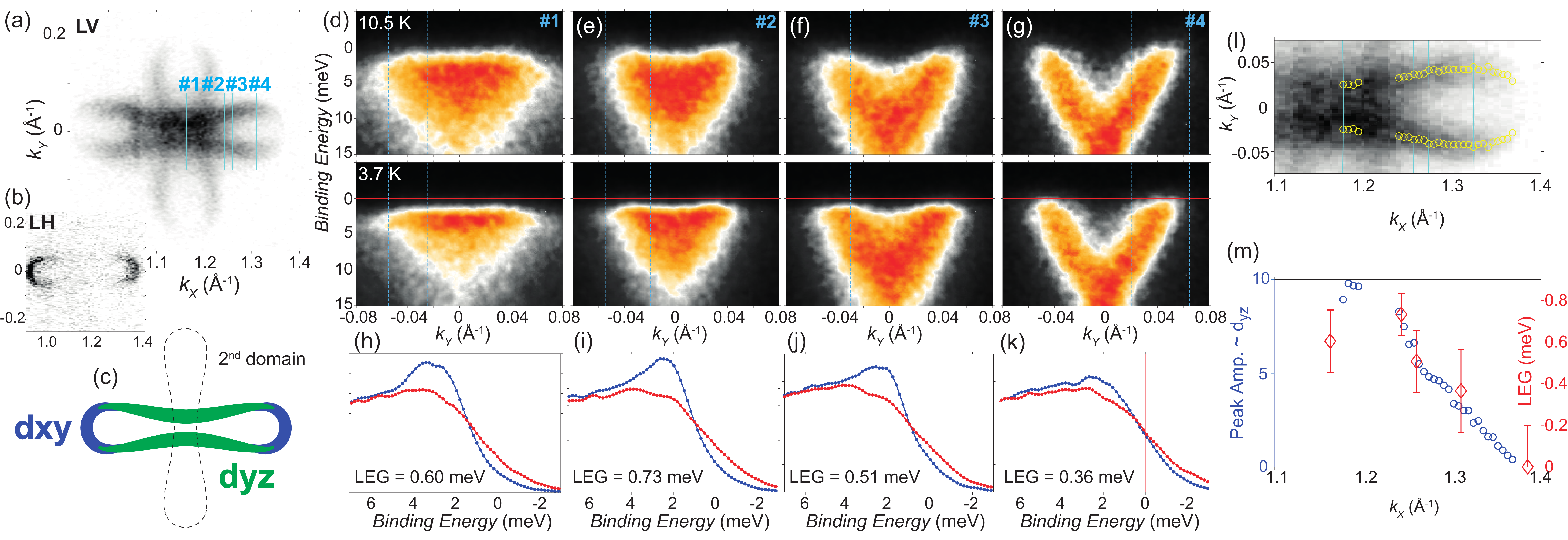}
	\caption{(a-b) Fermi surface maps of twinned FeSe at the A point (28~eV), schematically represented in (c). d-g) Cuts along $k_Y$, above and below $T_c$. Analysis of the EDCs in (h-k) reveals a SC gap that tends to decrease further away from the center of the peanut. l) Section of the Fermi surface used for peak-fitting analysis yellow circles indicate the peak positions. m) Peak amplitude from fitting the Fermi surface, which is proportional to the $d_{yz}$ orbital character. The decrease in the LEG with $k_X$ correlates with the decrease in $d_{yz}$ orbital character.}
	\label{fig:fig2}
\end{figure*}

By comparing ARPES spectra in the normal state and in the SC state at 3.7 K, we can determine the SC gap, $\Delta_{\mathbf{k}}$, around the Fermi surface. On bands where $\Delta_{\mathbf{k}}$ is finite, one should observe a shift of the ``leading edge" to higher binding energies compared with the Fermi-Dirac cut-off in the normal state due to the lack of single particle excitations within the gap.  Additionally one will observe a coherence peak at an energy scale close to $\Delta_{\mathbf{k}}$. Here we use the leading-edge gap (LEG) as a well-defined criterion of the superconducting gap; in the regime where the gap size is comparable to or less than the experimental resolution, the LEG is more reliable than the position of the coherence peak, though in general the LEG will underestimate the `true' gap due to the finite resolution~\cite{Evtushinsky2009}.  

The high-symmetry cuts in Fig.~\ref{fig:fig1}(c,d) reveal that the SC gap on this elliptical hole pocket is extremely anisotropic. The outer band dispersion in Fig.~\ref{fig:fig1}(c) does not show any significant difference through $T_c$, and the energy distribution curves (EDC) integrated in a narrow range around $k_F$ in Fig.~\ref{fig:fig1}(g) simply reveal a slightly sharper Fermi cut-off at low temperatures; within our resolution, the band is not gapped, though we cannot exclude a very small gap ($\lesssim$0.3 meV). By contrast, the inner band, seen weakly in LH polarization in Fig.~\ref{fig:fig1}(c) but which dominates in LV polarization in Fig.~\ref{fig:fig1}(d), is observed to bend back from the Fermi level in the SC state, forming a Bogoliubov band, which can be traced all the way across the pocket to the opposite $k_F$. The observation of fused Bogoliubov dispersions is possible only in a regime where $\Delta$ is comparable to $E_F$ \cite{Evtushinsky2011_arxiv}, as is the case for this particular hole band ($E_F \sim$ 10 meV). A coherence peak at 3.2 meV is observed, with a LEG of 1.6 meV; given the resolution of the measurement, the `true gap' at 3.7~K is estimated to be 2.1 meV (SI). 

This clear anisotropy, taken together with the reduced LEG seen in the data in Fig.~\ref{fig:fig1}(i,j) indicates that the LEG is maximal on the minor axis on the ellipse, but shrinks towards zero around the major axis. Thus the gap evolves monotonically around the Fermi surface, and in particular the gap tracks the orbital weight of $d_{yz}$ character, shown schematically in Fig.~\ref{fig:fig1}(f). In Fig.~\ref{fig:fig1} we only show data at the $\Gamma$ point, but qualitatively the angular dependence of the gap is similar at the Z point, except for a reduced magnitude (SI). 
Thus even though the $d_{yz}$ orbital weight is in the minority on the hole pocket, it dictates the symmetry of the gap. This is highly unusual: given the dominance of $d_{xz}$ weight, one might expect a much more uniform gap derived from the pairing of $d_{xz}$ states, but experimentally we see a highly anisotropic gap which is in antiphase with the $d_{xz}$ weight but correlates with the minority $d_{yz}$ weight. This is because there is a separate reservoir of $d_{yz}$ states to pair with on the electron pocket - hinting at the importance of inter-band pairing. 

In Fig.~\ref{fig:fig2}(a-c) we present the experimental and schematic electron pockets, here measured at 28 eV at the A point where the pockets are largest. The understanding of ARPES spectra of the electron pockets has been highly disputed \cite{Coldea2018Review, Watson2017d, Fedorov2016}. However recent high-resolution ARPES measurements on detwinned samples have shown that in one domain only the peanut shaped electron pocket oriented along the $a$ axis is detected \cite{Watson2017d}. Thus the Fermi surface map in Fig.~\ref{fig:fig2}(a) observes just two peanuts, one from each nematic domain in the twinned sample. The full implication of this one electron pocket structure will be discussed later, but we note in the schematic diagram in Fig.~\ref{fig:fig2}(c) that the observed peanut has largely $d_{yz}$ orbital character which couples to LV polarization in this geometry, while the tips of the peanut have $d_{xy}$ character, which can be detected in LH polarization. 

The electron pocket not only has a complex shape but also has a strongly anisotropic gap. The high-symmetry cut across the A point in Fig.~\ref{fig:fig2}(d,h) shows a LEG of 0.60 meV; a factor of 1.5-2 smaller than the maximum LEG at $\Gamma$. Moving away from the high symmetry point, in Fig.~\ref{fig:fig2}(e,i) the LEG initially becomes slightly larger but beyond that the general trend is for the gap to reduce as a function of $k_X$. Cut \#{}4 (Fig. \ref{fig:fig2}(g,k)) corresponds to where the band is largest, and also where the Fermi velocity is maximal. Here the gap becomes very small. At the tips of the pocket with $d_{xy}$ character we do not observe any gap (SI). We have additionally confirmed that the center of the pocket is gapped with a similar magnitude at the M point (SI). 

Crucially, the reduction in the size of the gap correlates with a reduced intensity in the Fermi surface map, as a function of $k_X$ away from the A point. In Fig.~\ref{fig:fig2}(m) we present the peak amplitude as a function of $k_X$, derived from fitting the Fermi surface map in Fig.~\ref{fig:fig2}(l). Assuming a constant matrix element, this quantity is proportional to the $d_{yz}$ orbital character on this band. Thus in this geometry we have a rather direct probe of $d_{yz}$ orbital weight, which is found to decrease continuously along the length of the peanut. We can therefore directly show that the magnitude of the LEG, superposed on Fig.~\ref{fig:fig2}(m), correlates very well with the $d_{yz}$ orbital weight \footnote{It is worth noting that, even in the case that the pairing interaction were momentum- and orbital-independent, some gap anisotropy would be induced due to density-of-states factors associated with the complex and anisotropic Fermi surfaces; one would typically expect a larger gap in regions with smaller density of states, i.e. larger Fermi velocity. However on the electron pockets we find that the gap is \textit{smallest} on the sections with highest Fermi velocity. This further substantiates our conclusion that the gap primarily follows the orbital character of the bands, with the details of the band dispersions only playing a secondary role.} .

We now focus on solving the linearized gap equation in FeSe for spin-fluctuation mediated pairing \cite{Hirschfeld2011,Scalapino2012}. In this approach it is important to not only accurately model the dispersion of the states present near the Fermi level, but also the orbital character of those states. Our model is based on a renormalized tight-binding parametrization of the electronic structure of FeSe in the tetragonal phase \cite{Rhodes2017}. In order to adapt this for the nematic state of FeSe, we require three key components: 1) a nematic order parameter which correctly describes the symmetry and energy separations of the bands at low temperatures, 2) the inclusion of spin orbit coupling which mixes the $d_{xz}$ and $d_{yz}$ orbital characters at the hole pocket, and most critically,  3) the inclusion of only the one experimentally observed electron pocket in the pairing calculations. 

\begin{figure}[h!]
	\includegraphics[width = 0.5\textwidth]{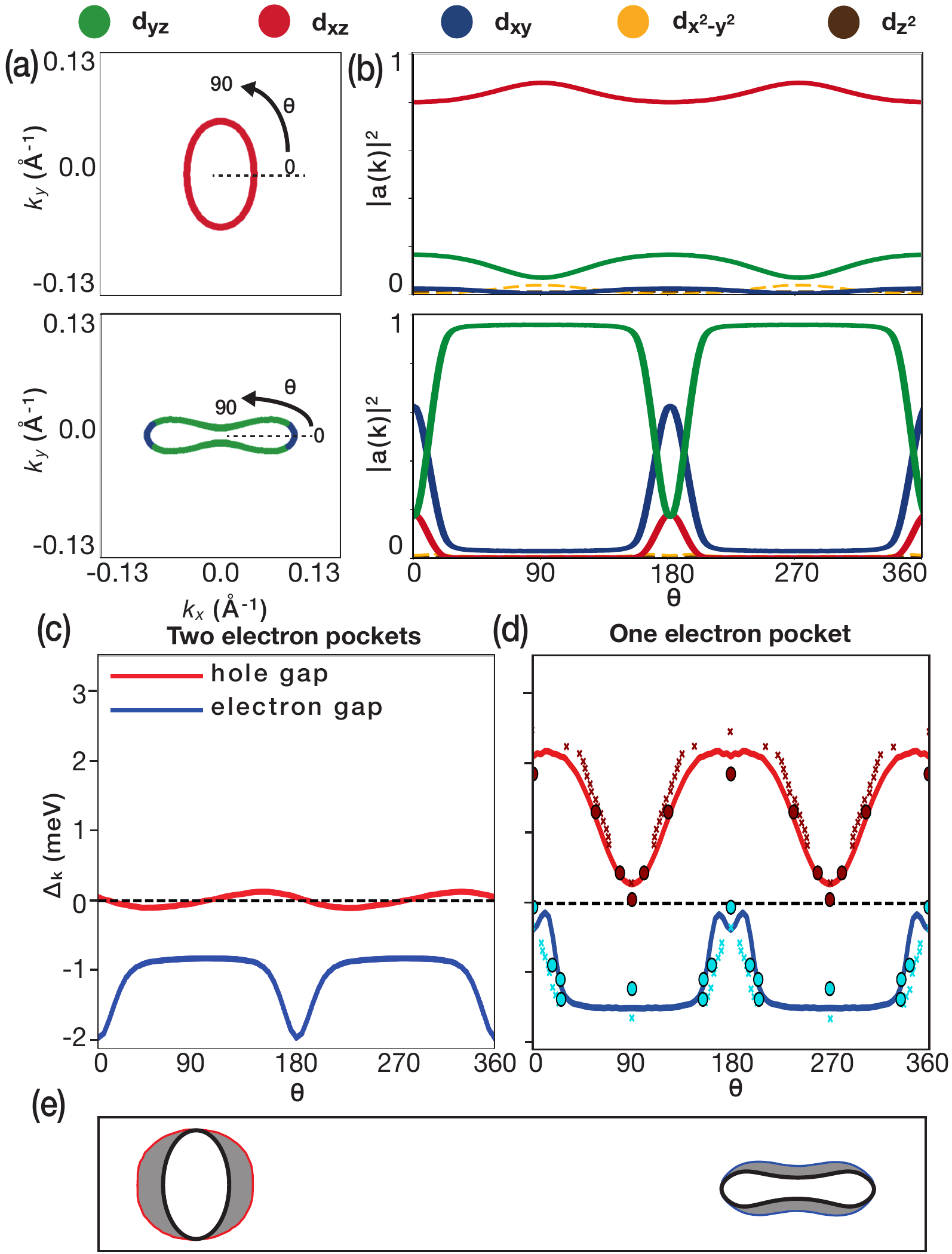}
	\caption{\label{fig:fig3}a) Close up of the hole pocket (top) and electron pocket (bottom) at $k_z = 0$. The color code describes the maximum orbital character and the axes are defined relative to the center of the pockets with $(0,0)$ for the hole pocket and $(\pi,0)$ for the electron pocket. b) Orbital characters as a function of angle for the hole pocket (top) and electron pocket (bottom). c) Momentum dependence of the gap structure for one hole pocket and two electron pockets with $U = 0.3$~eV. d) The same as c) but repeated without the electron pocket located at $(0,\pi)$. The colored circles correspond to the ARPES data reported here whilst the crosses are taken from quasiparticle interference experiments \cite{Sprau2017}. e) Representation of the results of (d) around the Fermi surface. }
\end{figure}

In Fig. \ref{fig:fig3}(a) we plot the model Fermi surface in the $k_z = 0$ plane. These pockets correctly describe the shape, size and curvature of the low temperature Fermi surface observed by ARPES \cite{Watson2015,Watson2016,Watson2017d,Fedorov2016}, with a smoothly elliptical hole pocket and a peanut-shaped electron pocket; for details of the nematic order parameter, see SI. In Fig.~\ref{fig:fig3}(b) we present the corresponding orbital characters. For the hole pocket, the orbital character is predominantly of $d_{xz}$ weight, however there is a contribution of the $d_{yz}$ weight which reaches 16.5\% of the total weight along the minor axis in agreement with the experimental determination from Fig. \ref{fig:fig1}(e). For the electron pocket at $(\pi,0)$, the dominant orbital character is $d_{yz}$ with the ends of the peanut being mostly $d_{xy}$. 

It can be seen in Fig. \ref{fig:fig3}(c) that when all states are considered in the linearized gap equation, including the expected electron pocket at $(0,\pi)$ the gap structure does not reproduce the experimental results. The gap on the hole pocket is predicted to have nodes and be much smaller in magnitude than the gap at the $(\pi,0)$ electron pocket, which broadly follows the $d_{xy}$ orbital character. However when we remove the contribution to SC pairing from the electron pocket at $(0,\pi)$, we obtain a highly anisotropic gap structure very close to the experimental results, as presented in Fig. \ref{fig:fig3}(d). Here the hole pocket is nodeless but with a strong anisotropy which follows the minority $d_{yz}$ orbital character shown in Fig. \ref{fig:fig3}(b). The gap function at the electron pocket is also nodeless, with a reduced magnitude and opposite sign to the hole pocket, and closely follows the $d_{yz}$ orbital weight \footnote{In these results, the normalized eigenvector $g(\mathbf{k})$ has been scaled by a single constant to best fit the experimental gap values as observed in this work and by Bogoliubov quasiparticle interference experiments.}. The success of this calculation shows that as long as one carefully accounts for the details of the experimental Fermi surface, including spin-orbit coupling, nematic order, and considering only the one observed electron pocket, this gap structure which follows the $d_{yz}$ weight with a sign-change between the pockets is naturally the leading instability, within spin-fluctuation pairing theory. 
Thus we have provided an alternative explanation for the anisotropy of the gap structure, which does not require any artificial suppression of quasiparticle weights of certain orbitals \cite{Kreisel2017a}. Instead the gap is dictated mainly by the topology and orbital character of the Fermi surface, with only the $d_{yz}$ states present on both hole and electron pockets. 

Our results can be considered as an independent verification of the gap structure determined by Bogoliubov quasiparticle interference measurements \cite{Sprau2017} and specific heat experiments \cite{Sun2017}. Previously Xu et. al. \cite{Xu2016} found the gap anisotropy at the Z point in FeSe$_{0.93}$S$_{0.07}$, but here we have extended this to measure the gap on the electron pockets, as well as showing the gap at $\Gamma$. Additionally, in recent months two groups have reported the SC gap structure on the hole pocket of FeSe using laser-ARPES: Liu et. al. reported a similar gap anisotropy to us \cite{Liu2018_arxiv}, while some details vary in Hashimoto et. al. \cite{Hashimoto2018}. 

In summary, we have measured the full gap structure and have shown both experimentally and theoretically a direct link between the $d_{yz}$ orbital content and the gap magnitude. 
Whilst the relationship between the  $d_{yz}$ orbital and the gap is theoretically complex, the fact that such a link exists provides strong evidence for spin fluctuation mediated superconductivity in FeSe. 
More precisely, it is evidence that the pairing interactions derive from the local, instantaneous and repulsive Coulomb interactions, in sharp contrast to the retarded, attractive and orbitally-agnostic electron-phonon pairing. 
Finally, our results also emphasize the impact of an incoherent electron pocket in FeSe.

{\it Note added:} While we were preparing our manuscript, a preprint by Kushnirenko \textit{et al.} \cite{Kushnirenko2018_arxiv} appeared, who also report the gap anisotropy on both hole and electron pockets with synchrotron ARPES.

\begin{acknowledgments}
	\section{acknowledgments}
	We thank P.~King, M.~Hoesch, A.~V.~Chubukov, R.~M.~Fernandes and A.~I.~Coldea for useful discussions. We thank Diamond Light Source for access to Beamline I05 (Proposal No. SI17532) that contributed to the results presented here. 
\end{acknowledgments}



%

\clearpage

\onecolumngrid
\appendix
 
 \section{Supplimental Material}
 \subsection{Estimates of orbital character at the $\Gamma$ point}
 \begin{figure}[h]
 	\centering
 	\includegraphics[width=0.7\linewidth]{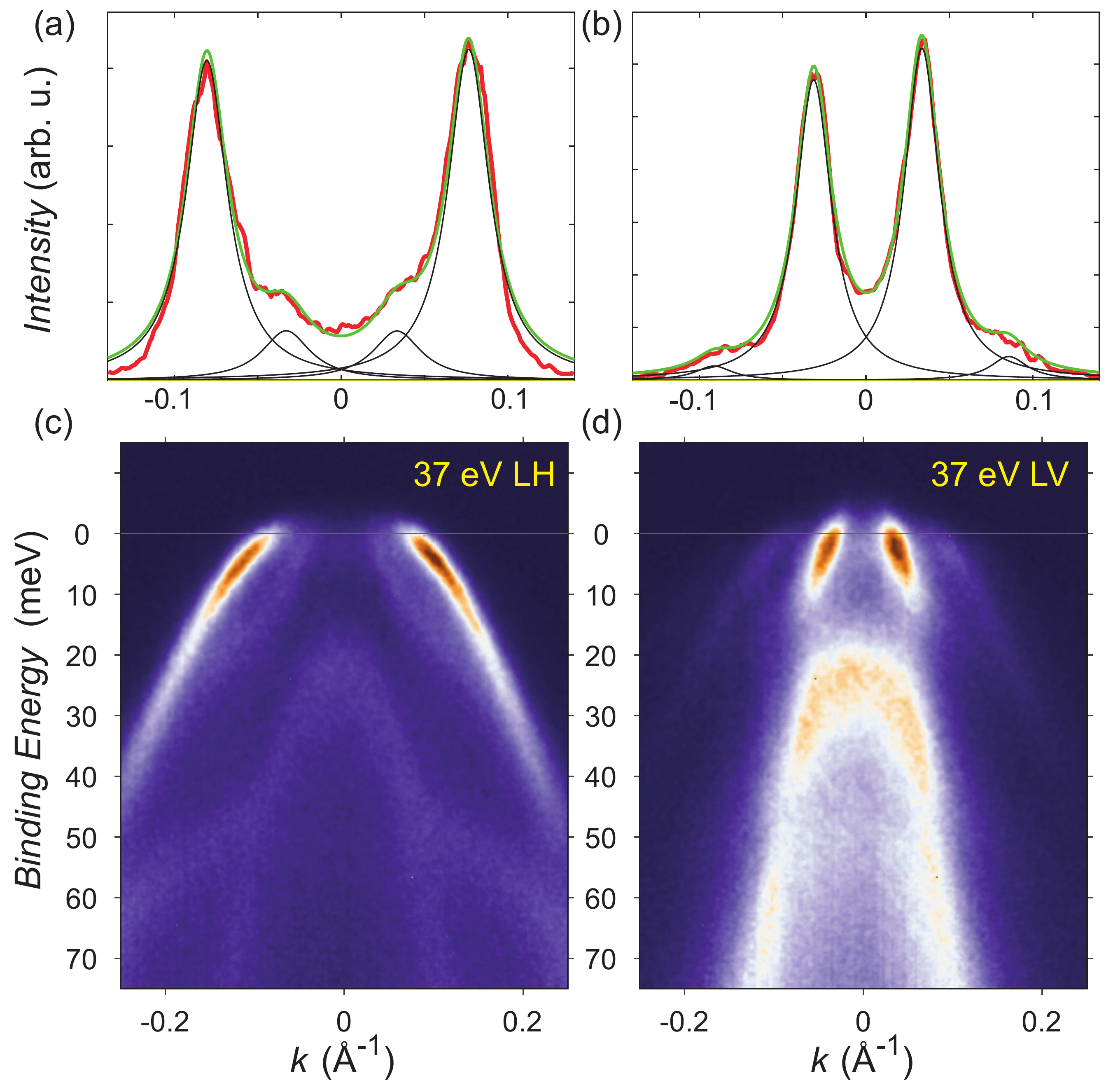}
 	\caption{High symmetry dispersions and MDCs at the Fermi level for a cut through the hole pocket at the $\Gamma$ point at 10K in LH and LV; (c) is identical to Fig.~1(c) of the main text but plotted on a wider scale. Green lines and black lineshapes indicate the 4-Lorentzian fit to the data. The sample is twinned, but in each polarization it appears almost detwinned; it is necessary to have excellent angular resolution to distinguish the second, much weaker band from the other domain.}
 	\label{figS1}
 \end{figure}
 
 Qualitative statements on the orbital characters of a band can be made on the basis of the ARPES matrix elements, but quantitative analysis is not straightforward. We have shown in the main text that the hole pocket at $\Gamma$ must be mostly composed of $d_{xz}$ orbital character in order to explain the switching behavior with different polarizations. It cannot purely be $d_{xz}$, however, since traces of the other ellipse can be observed in the high-symmetry cuts, shown in Fig.~\ref{figS1}(c,d). In order to extract some numerical estimates of orbital characters, we fit the MDCs in Fig.~\ref{figS1}(a,b) with pairs of Lorentzians. From the amplitude of the peaks, the ratio of the Inner ($I$) to the Outer ($O$) peaks in the LV data gives us $I_{xz}/O_{yz} \approx 16$ and in the LH data $I_{yz}/O_{xz} \approx 0.19$. Here we have assumed that the matrix elements are perfect, i.e. that the LV couples 100\% to the $d_{YZ}$ orbital and 0\% to the $d_{XZ}$ orbital and that there are only $d_{xz}$ and $d_{yz}$ contributions to the bands. We additionally must assume that the volume fractions of each domain are exactly equal, and also that the polarization of the incident light is perfect (with respect to the sample surface). If we further impose that $I_{xz}+I_{yz} = 1$, likewise for $O$, we have linear simultaneous equations which can be solved to estimate that the $d_{yz}$ = 0.18(5) on the minor axis and = 0.05(2) on the major axis. The error bars here are estimated from the range of fitting parameters which would also adequately describe the data, as there is some flexibility in the fit. The numbers obtained are consistent with the qualitative observation that the pocket must be mostly $d_{xz}$, as well as the idea that the $d_{yz}$ weight should be maximal on the shorter axis of the ellipse; that said, there are some weak points to this analysis and the numbers should only be taken as a rough estimate. 
 Note that this estimate solely applies to the $\Gamma$ point; unfortunately for the 23 eV Z point we have not collected data in LV polarization for technical reasons, but qualitatively there should be more $d_{yz}$ weight around the Z point as the pocket is much larger so the orbital characters deviate less from the tetragonal phase.

 \subsection{Comment on orbital-selective coherence}
 FeSe is a strongly correlated system in which the quasiparticle $Z$ factors at $E_F$ are less than unity, and could in principle be orbitally-differentiated. Recent work has hypothesized that the tiny superconducting gaps on Fermi surface segments with mainly $d_{xz}$ or $d_{xy}$ character is a consequence of these orbitals being particularly incoherent \cite{Sprau2017,Kreisel2017a,Kostin2018_arxiv}. In those works, it was suggested that the Z factor for $d_{yz}$ remains close to unity, whereas for $d_{xz}$ and $d_{xy}$ it is small. However we have shown via the analysis of the polarization switching effect that at low temperatures the $d_{xz}$ orbital weight dominates the hole pocket. Moreover the ends of the electron pocket with $d_{xy}$ orbital character can be observed in the right conditions (LH polarization) as shown in Fig.~2(b) of the main text. Both bands are as sharp and coherent as $d_{yz}$ sections, they simply have vanishing gaps. Thus we find that the idea of having significant reduction of the quasiparticle weights of the $d_{xz}$ and $d_{xy}$ orbitals compared with the $d_{yz}$ orbital \cite{Sprau2017,Kreisel2017a,Kostin2018_arxiv} is incompatible with spectroscopy of the normal state single particle spectral function.   
 
 We note that the motivation for introducing the orbitally-selective Z factors was the observation that you somehow need to dramatically break fourfold symmetry in the system to obtain the highly anisotropic experimental gap structure, particularly on the hole pocket - simply tweaking the band structure with some nematic order parameter is insufficient. However we contend that the required source of anisotropy in the normal state is the one-peanut effect, i.e. a \textit{pocket-}specific loss of coherence, not an \textit{orbital-}selective effect.

 \clearpage

 \subsection{Gap structure at the Z point}
 
 \begin{figure}[h]
 	\centering
 	\includegraphics[width=0.95\linewidth]{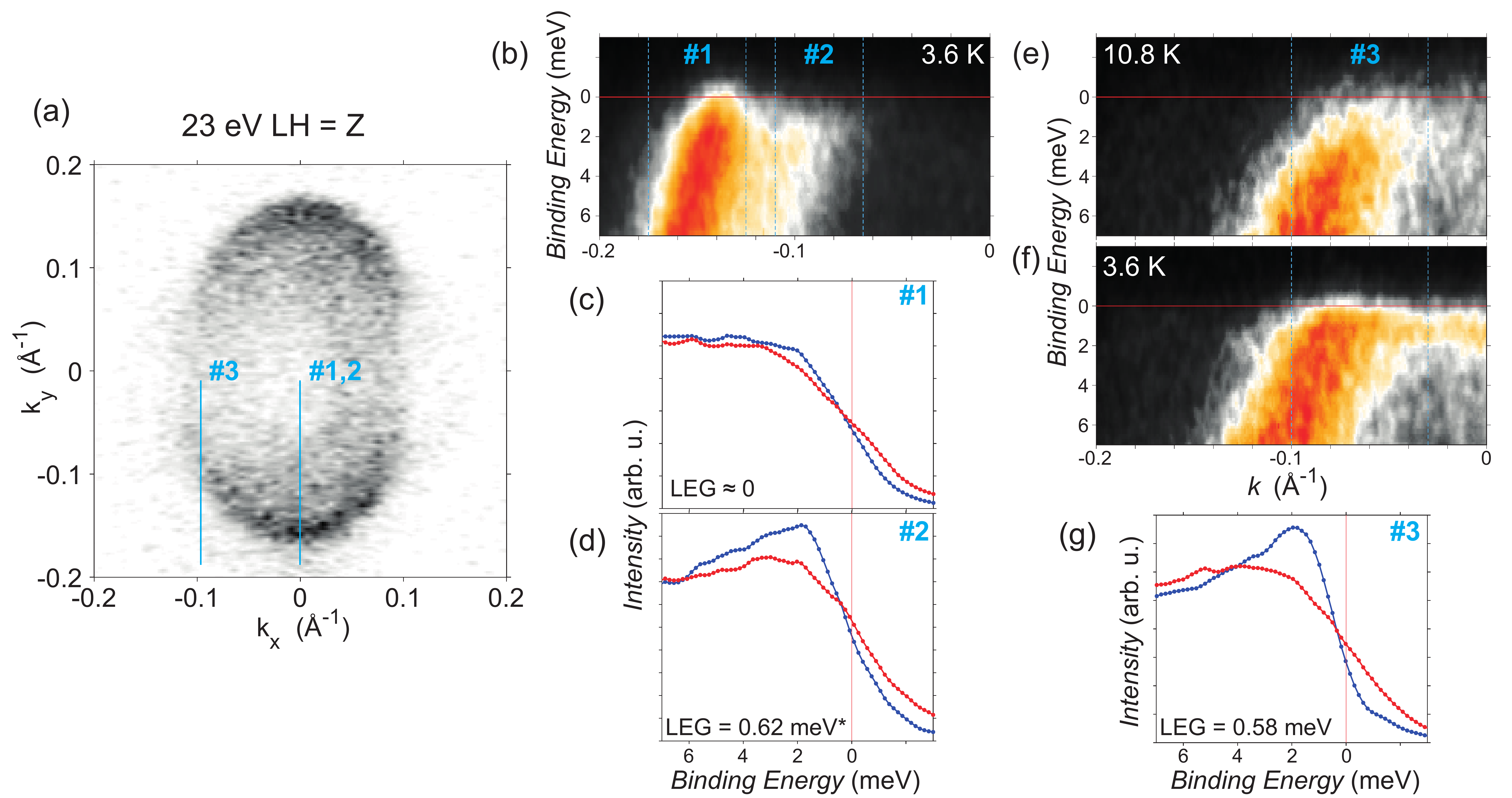}
 	\caption{(a) Fermi surface of the hole pocket of twinned FeSe. (b) High symmetry cut, focusing on the hole band dispersions at $k_y < 0$; the EDCs in (c) and (d) reveal that the weaker inner band is gapped, whereas the outer band is ungapped. Note that the value of the LEG in (d) may be a slight underestimate due to the leaking spectral weight of the brighter and ungapped outer band nearby. (e,f) Off-high-symmetry band dispersion, showing the development of a clear Bogoliubov branch below $T_c$. (g) The EDCs reveal a substantial LEG and coherence peak; a hint of the coherence peak in the unoccupied states is also detected in this geometry.}
 	\label{figS2}
 \end{figure}
 
 In the main text we show hole pocket data only at 37 eV, the $\Gamma$ point. However there is a significant $k_z$ dispersion and the pocket is much larger at Z, where it will also have different orbital composition; the $d_{xz}$ character will still have the largest contribution but is less dominant, especially on the minor axis of the ellipse, where we expect a higher percentage of $d_{yz}$ weight than the 18\% estimated at the $\Gamma$ point. Our data in Fig.~\ref{figS2} shows that, despite these variations, the overall gap structure at Z is very similar to $\Gamma$: Fig.~\ref{figS2}(c) shows no measurable gap on the outer band, but the inner band from the minor axis of the ellipse, only weakly appearing in this geometry, is gapped. 
 
 We find that the magnitude of the gap at Z is smaller than at $\Gamma$; on this point, we differ from the recent study of Ref.~\cite{Kushnirenko2018_arxiv}, and the earlier study of Ref.~\cite{Xu2016} on FeSe$_{0.93}$S$_{0.07}$, where much smaller or no gaps were detected at the $\Gamma$ point. In this study we find that the gap on the minor axis of the hole pocket at $\Gamma$ is the largest found anywhere on the 3D Fermi surface. 
 
 \clearpage
 
 \subsection{Gap structure at the M point}
 \begin{figure}[h]
 	\centering
 	\includegraphics[width=0.85\linewidth]{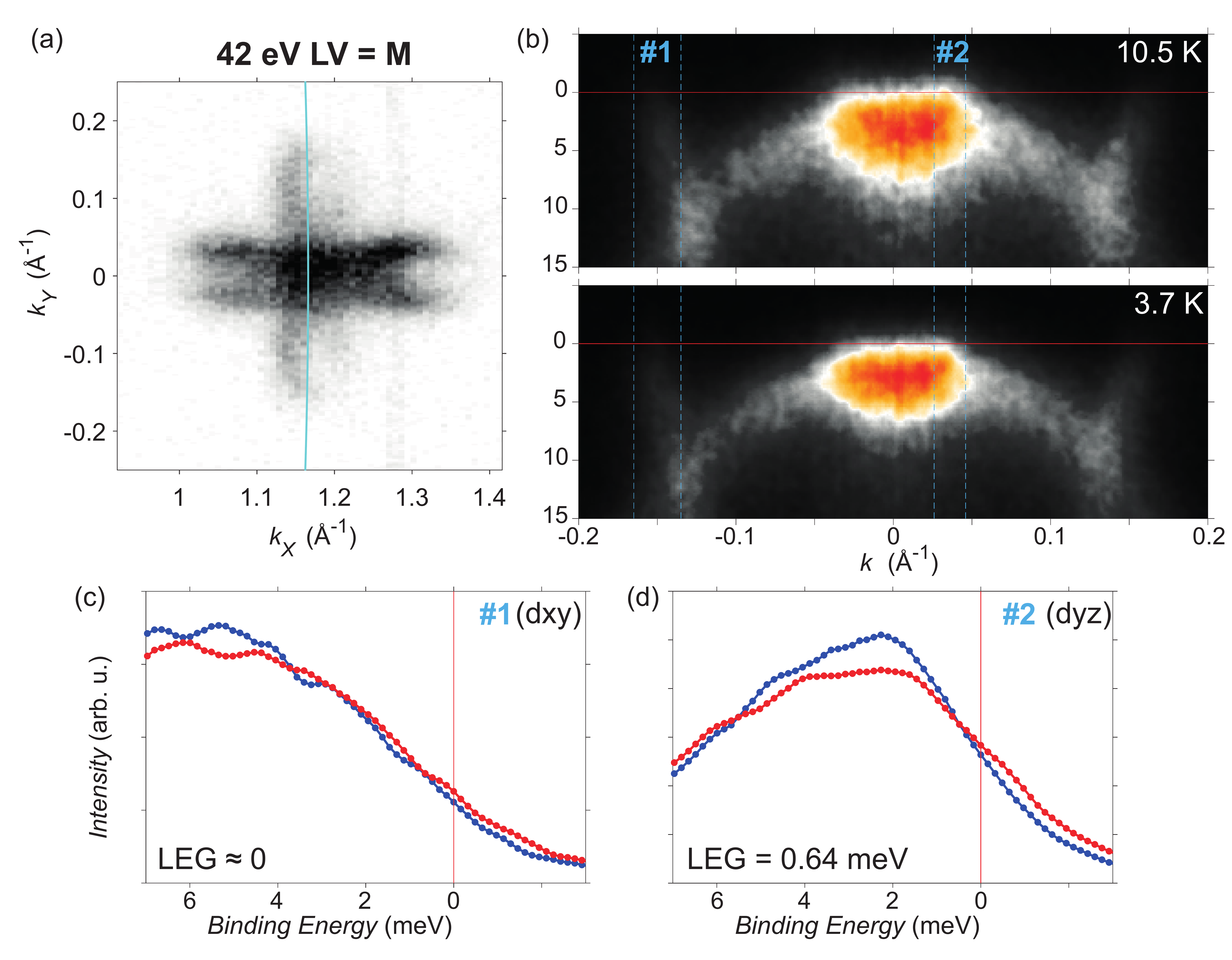}
 	\caption{(a) Fermi surface of FeSe at the M point (42 eV, LV), where the pocket has a smaller size but otherwise similar structure to the A point presented in the main text. (b) Dispersions above and below $T_c$; note that the cut is very slightly off the high symmetry axis in order to obtain higher intensity on the outer band with $d_{xy}$ character; the matrix elements for $d_{xy}$ orbitals are complex. (c) EDCs for the outer band with $d_{xy}$ character, showing no gap, and (d) for the inner band with $d_{yz}$ character, which shows a finite leading edge gap and a coherence peak, indicating the presence of a gap on this pocket. Note that the detection of $\sim$meV-scale superconducting gaps at 42 eV photon energy is  challenging and the magnitude of the LEG should be taken with some caution; nevertheless we are confident in the result that the pocket is also gapped at M.}
 	\label{fig:suppfigmpoint}
 \end{figure}
 
 \clearpage
 
 \subsection{Simulating $\Delta$ from LEG}
 
 \begin{figure}[h]
 	\centering
 	\includegraphics[width=0.9\linewidth]{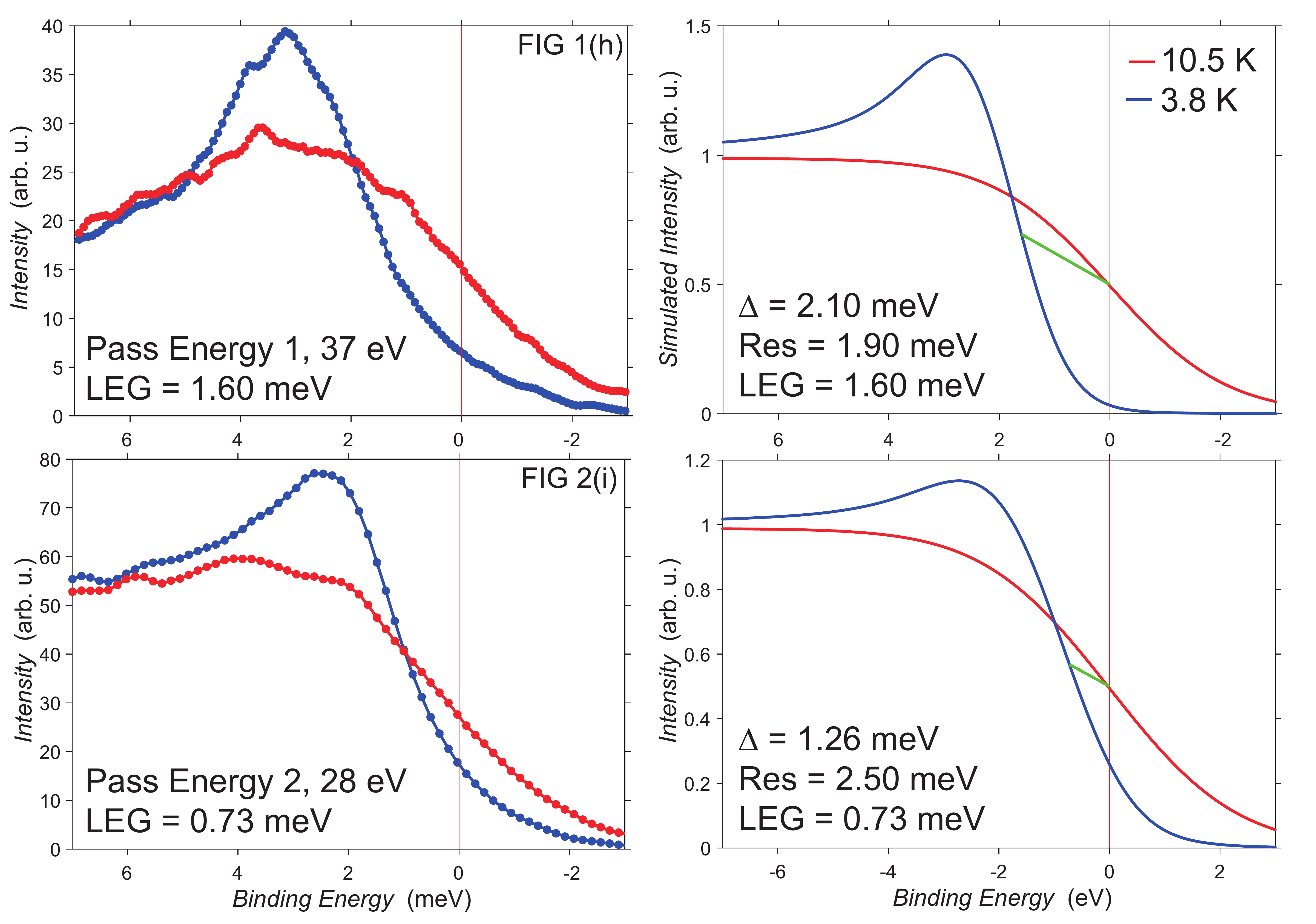}
 	\caption{Comparison of experimental EDCs with simulations of the Fermi-Dynes function at our experimental temperatures above and below $T_c$, at a realistic experimental resolution. From such simulations we can infer a "true gap" $\Delta$ from the leading edge gap and the experimental resolution. Note that this is still the gap at 3.8~K, not the zero temperature limit. For the purposes of comparison with our calculations, we use these values $\Delta$ in Fig.~3 of the main text.}
 	\label{fig:suppfigsimulatedeltafromleg}
 \end{figure}
 
 \clearpage
 
 \subsection{Tight binding model}
 The tight binding model used in the present calculations was introduced in Ref. \cite{Rhodes2017}, where the hopping parameters were fitted to experimental band dispersions derived from ARPES spectra at 100~K. In this work we keep the same parameters except that we have modified the value of the parameter $\epsilon_2$ from -0.05050~eV to +0.02~eV. This removes $d_{x^2-y^2}$ weight from the hole bands, which we believe to be artificial. 
 We have used a spin orbit coupling value of $\lambda_{SOC} = 19$~meV to accurately describe the experimental splitting between the $d_{xz}$ and $d_{yz}$ bands at the hole pocket \cite{Watson2017}. The form of spin orbit coupling is discussed in Ref. \cite{Saito2015a}.
 
 The chemical potential in this model is also very sensitive to changes in temperature and shifts to the bands \cite{Rhodes2017}. For this reason, we self consistently calculate the change to the chemical potential by including the nematic order parameter presented below with a strength of $\Delta_{nem} = 29$~meV using the method described in \cite{Rhodes2017} and determine that the chemical potential should increase from 12~meV at 100~K to 20~meV at 10~K. We thus use a value of $\mu=20$~meV in these calculations. This chemical potential shift also accounts for the observed momentum independent decrease of the $d_{xy}$ band position \cite{Watson2016,Watson2017}.
 
 \subsection{Nematic order parameter}
 
 We now discuss the choice of the nematic order parameter. In Ref. \cite{Watson2016} we introduced a `unidirectional nematic bond order' parameter to describe the observed symmetries of the band shifts determined from ARPES studies \cite{Watson2016,Fedorov2016}. 
 
 \begin{equation}
 \label{unidirectional_nematic}
 h = \Delta_{nem}(n'_{yz}-n'_{xz})\cos(k_x).
 \end{equation}
 
 Here $h$ is the nematic order parameter, $n'_{xz/yz}$ describe the hopping between the $d_{xz/yz}$ orbital on the first and second Fe atom in a two Fe unit cell, making this a bond-centered order parameter. $\Delta_{nem}$ is the magnitude of the nematic ordering.
 
 Here we consider a generalised form of this order, expanding eq. \eqref{unidirectional_nematic} into an extended s-wave term and a symmetry-allowed term,
 
 \begin{equation}
 \begin{split}
 h = &\frac{\Delta_h}{2}(n'_{yz}-n'_{xz})(\cos(k_x) + cos(k_y))\\
 +&\frac{\Delta_e}{2}(n'_{yz}-n'_{xz})(\cos(k_x) - cos(k_y)).
 \end{split}
 \end{equation}
 
 The first term is the extended s-wave bond order, which gives rise to the additional band splitting and the elliptical distortion of the hole pocket, while having no effect at the electron pockets. Note that with this definition, the sum of the cosine terms effectively gives a factor of two at the $\Gamma$ point, such that the low temperature splitting is $\approx \sqrt{(\lambda_{SO}^2+(2\Delta_h)^2)}$. The second term affects only the electron bands, and is a symmetry-allowed hopping, which nevertheless in our model will onset at $T_s$ \cite{Watson2016} and be linked to nematic order. This term causes both of the electron pockets to distort into `peanuts'. This expanded form has the benefit that it allows us to change the magnitude of the nematic shifts at the hole pocket and electron pocket independently, whilst still retaining the symmetries of the band shifts. In the case that $\Delta_h = \Delta_e$ we reobtain equation \eqref{unidirectional_nematic}.

\begin{figure}[h!]
 	\includegraphics[width = 0.9\textwidth]{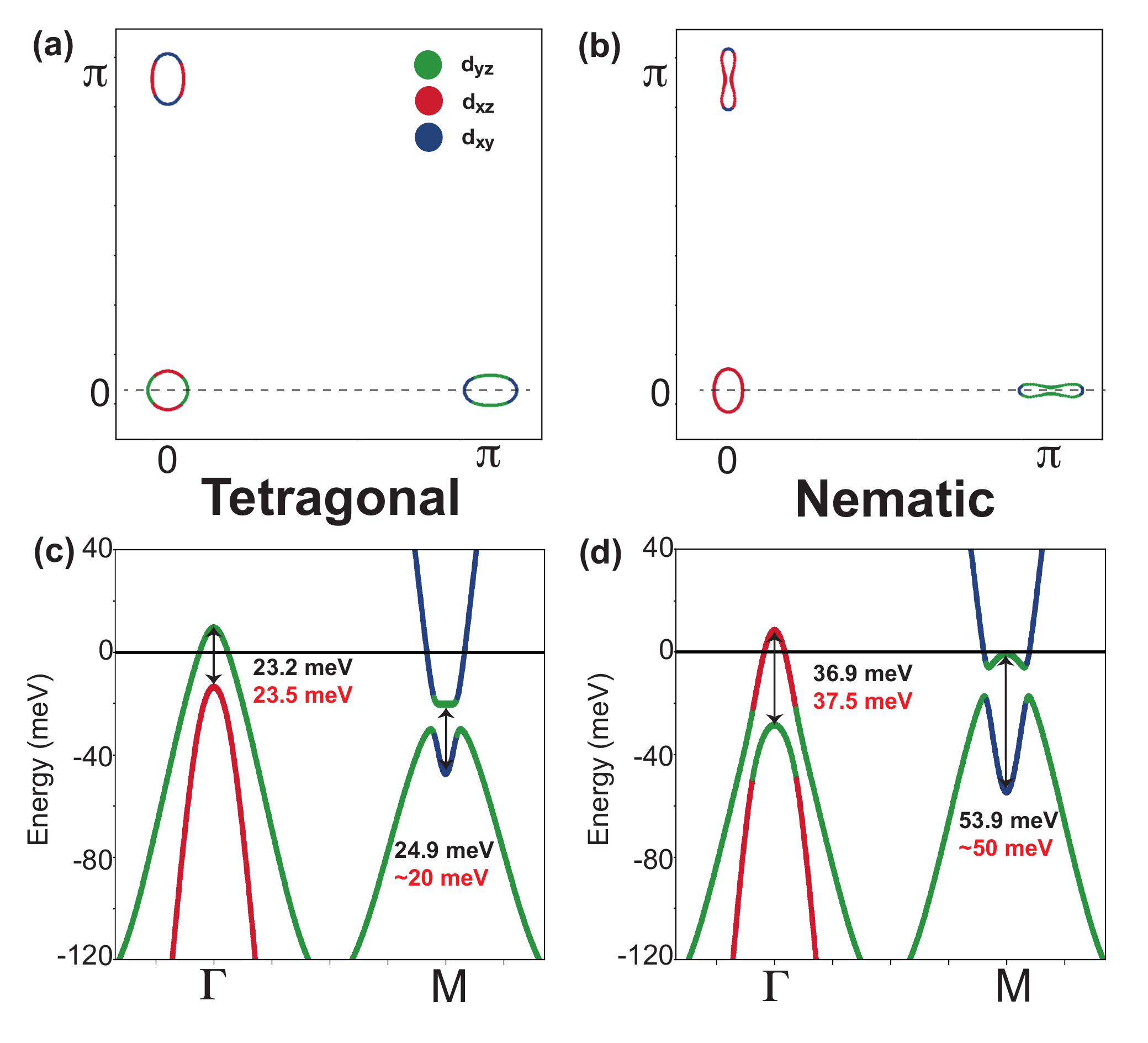}
 	\caption{\label{Nematic_Shifts} Comparison of the tight binding model in the tetragonal phase and with the inclusion of the nematic order parameter. a,b) Fermi surface in the tetragonal and nematic phase for the $k_z = 0$ plane. c,d) corresponding band dispersions between $(0,0)$ and $(\pi,0)$ indicating the separations between the bands at the high symmetry points. black values are calculated separations, whilst red values are experimental values taken from ref. \cite{Watson2017,Watson2016}. In the tetragonal phase $\lambda_{SOC} = 19$~meV, $\Delta_{nem} = 0$~meV and $\mu = 12$~meV and in the nematic phase $\lambda_{SOC} = 19$~meV, $\Delta_{nem} = 29$~meV and $\mu = 20$~meV.}
 \end{figure}

 To reproduce the quantitative band positions determined from detailed temperature dependent ARPES studies \cite{Watson2017,Watson2016}, we set $\Delta_e = 29$~meV and $\Delta_h = \frac{1}{2}\Delta_e = 14.5$~meV, and the chemical potential is self consistently adjusted. We compare the distortions to the Fermi surface as well as the calculated band separations with the experimental values in Fig \ref{Nematic_Shifts}.

 Other groups have proposed different parameterizations of the low temperature Fermi surface of FeSe. Given that only one of the electron pockets is observed, the exact form of the nematic order is not fully constrained. However we suggest our formulation is advantageous since: (1) We find a smooth elliptical hole pocket, as found in experiment, but in contrast to the pinched structure found in Ref. \cite{Kreisel2017a} and elsewhere; (2) the relative weight of $d_{xz}$ and $d_{yz}$ orbital characters agrees with the experimental estimate; (3) by construction, our model also correctly reproduces the tetragonal phase, if the nematic order is switched off and chemical potential adjusted.  
 
 \subsection{Linearized gap equation in the presence of spin orbit coupling}
 Here we discuss the linearized gap equation used to calculate the results of Fig. 3c and 3d in the main text.
 
 The linearized gap equation for a multiorbital tight binding model  \cite{Graser2009} is
 \begin{equation}
 \label{eigenvalue equation}
 -\frac{1}{4\pi^2}\sum_{\nu}\oint_{C_\nu}\frac{d\mathbf{k'}_\parallel}{v_F(\mathbf{k'})}\Gamma^{SOC}_{\mu\nu}(\mathbf{k},\mathbf{k'}) g_\alpha(\mathbf{k'}) = \lambda_\alpha g_\alpha(\mathbf{k}).
 \end{equation}
 
 Where we perform an integral of the pairing vertex $\Gamma^{SOC}_{\mu\nu}(\mathbf{k},\mathbf{k'})$ over all $\mathbf{\mathbf{k'}}$ states on the Fermi surface for each band, $\nu$ . By solving this eigenvalue equation and taking the eigenvector, $g(\mathbf{k})$, corresponding to the leading eigenvalue, $\lambda$, we can determine both the leading pairing symmetry of the pairing vertex and the momentum dependence of that pairing. Here $v_F(\mathbf{k}) = \nabla E_\mu(\mathbf{k})$ is the Fermi velocity, which is the derivative of the eigenvalues $E_\mu(\mathbf{k})$ of the original Hamiltonian of the system at band $\mu$ and momentum $\mathbf{k}$ with respect to $\mathbf{k}$.

 In the presence of spin orbit coupling, the pairing vertex for the singlet state in the band basis takes the form presented in Ref. \cite{Saito2015a},
 
 \begin{equation}
 \Gamma^{SOC}_{\mu\nu}(\mathbf{k},\mathbf{k'}) = \Big[\Gamma^{\Uparrow\Downarrow\Uparrow\Downarrow}_{\mu\nu}(\mathbf{k},\mathbf{k'}) - \Gamma^{\Uparrow\Downarrow\Downarrow\Uparrow}_{\mu\nu}(\mathbf{k},\mathbf{k'})\Big],
 \end{equation}
 where
 \begin{equation}
 \Gamma^{\Sigma\bar{\Sigma}\Lambda\bar{\Lambda}}_{\mu\nu}(\mathbf{k},\mathbf{k'}) =  \sum_{stpq}\sum_{\sigma\bar{\sigma}\lambda\bar{\lambda}} a_{\mu\Sigma}^{t\sigma*}(\mathbf{k}) a_{\mu\bar{\Sigma}}^{s\bar{\sigma}*}(-\mathbf{k}) Re[\Gamma^{pq;\lambda\bar{\lambda}}_{st;\sigma\bar{\sigma}}(\mathbf{k},\mathbf{k'})] a_{\nu\bar{\Lambda}}^{p\bar{\lambda}}(\mathbf{-k'}) a_{\nu\Lambda}^{q\lambda}(\mathbf{k}').
 \end{equation}
 Here $a_{\mu\Sigma}^{t\sigma}(\mathbf{k})$ is the eigenvector of the original Hamiltonian in the presence of spin orbit coupling which connects the orbital  and spin basis ($s,p,q,t$ and $\sigma,\lambda$) with the band and pseudospin - band basis ($\mu,\nu$ and $\Sigma,\Lambda$). 
 
 The pairing vertex in orbital space \cite{Saito2015a} is then defined as 
 \begin{equation}
 \Gamma^{pq;\lambda\bar{\lambda}}_{st;\sigma\bar{\sigma}}(\mathbf{k},\mathbf{k'})  = V^{c}_{pq;st}\delta_{\sigma\lambda}\delta_{\bar{\sigma}\bar{\lambda}} + V^{s}_{pq;st}\mathbf{\vec{\sigma}}_{\sigma\lambda}\mathbf{\vec{\sigma}}_{\bar{\sigma}\bar{\lambda}}.
 \end{equation}
 \noindent $\mathbf{\vec{\sigma}}_{\sigma\lambda}$ being a vector of Pauli matrices. By performing the spin sum over these Pauli matrices we obtain a pairing vertex of the form
 \begin{equation}
 \Gamma^{pq;\lambda\bar{\lambda}}_{st;\sigma\bar{\sigma}}(\mathbf{k},\mathbf{k'})]  =
 \begin{cases}
 V^{c}_{pq;st} + V^{s}_{pq;st}, & \sigma = \lambda = \bar{\sigma} = \bar{\lambda} \\
 V^{c}_{pq;st} - V^{s}_{pq;st}, & \sigma = \lambda \neq \bar{\sigma} = \bar{\lambda} \\
 2V^{s}_{pq;st}, & \sigma = \bar{\lambda} \neq \lambda = \bar{\sigma}  \\
 0, & otherwise.
 \end{cases}
 \end{equation}
 
 Where $V^{c/s} = \frac{1}{2}U^{c/s}\chi^{c/s}U^{c/s}$, is the product of the RPA susceptibility matrix ($\chi^{c/s}$) in the charge (c) or spin (s) channel multiplied by the local interaction matrix ($U^{c/s}$), where   $\chi^{c/s} = \chi^0[1\pm U^{c/s}\chi^0]^{-1}$  and 
 
 \begin{equation}
 U^{s} =
 \begin{cases}
 U, & p= q = s = t \\
 U', & p = s \neq q = t \\
 J, & p = q \neq s = t \\
 J', & p = t \neq q = s \\
 0, & otherwise,
 \end{cases}
 \end{equation}
 
 \begin{equation}
 U^{c} =
 \begin{cases}
 U, & p = q = s = t \\
 -U'+2J, & p = s \neq q = t \\
 2U'-J, & p = q\neq s = t \\
 J', & p = t \neq q = s \\
 0, & otherwise.
 \end{cases}
 \end{equation}
 In these calculations we assume spin rotational invariance such that $U' = U - 2J$, and set $J = \frac{U}{6}$ with $J' = J$. U is set to 0.3~eV throughout.
 
 The non-interacting susceptibility matrix, $\chi^0$, is calculated from the polarization bubble
 \begin{equation}
 \label{ni_suscep}
 \chi^{0}_{pq;st}(\textbf{q},i\omega_m) = -\frac{1}{N\beta}\sum_{\textbf{k},i\omega_n} G_{sp}(\textbf{k},i\omega_n)G_{qt}(\textbf{k}+\textbf{q},i\omega_n + i\omega_m).
 \end{equation}
 Here, $\beta = \frac{1}{k_bT}$ where T is the temperature of the system, N is the number of Fe atoms, which is 1 in the present calculation. We calculate $\chi_0$ without the inclusion of spin orbit coupling as it has been shown that spin orbit coupling will only strongly affect the susceptibility when very close to a magnetic instability \cite{Saito2015a}.  The Greens function in the orbital basis is defined 
 \begin{equation}
 G_{sp}(\textbf{k},i\omega_n) = \sum_{\mu}\frac{a^s_\mu(\textbf{k})a^{p*}_\mu(\textbf{k})}{i\omega_n - E_\mu(\textbf{k})}
 \end{equation}
 Here, $\omega_n = (2n+1)\pi T$, is the Matsubara frequency. We have used a 32x32x16 $\mathbf{k}$-grid to determine equation \eqref{ni_suscep}, and set the matsubara cutoff to $n_{cutoff} = 128$ for a temperature of 10~K.
 
 For the purposes of our susceptibility calculation, we include contributions from all states in the model, including the second electron pocket which is not detected experimentally. Thus the comparison presented in Fig. 3 of the main text shows the effect of simply removing the $\mathbf{k}$-states associated with the $(0,\pi)$ electron pocket. The gap structure we obtain is largely independent of the details of the susceptibility (and the values of $U$ and $J$) and is primarily dictated by the details of the distribution of orbital weight at the Fermi level. The success of this calculation suggests that we capture the essential physics of spin-fluctuation mediated superconductivity simply within the RPA approximation.

 \subsection{Comparison of the gap structure without spin orbit coupling}
 
 \begin{figure}[h!]
 	\includegraphics[width = \textwidth]{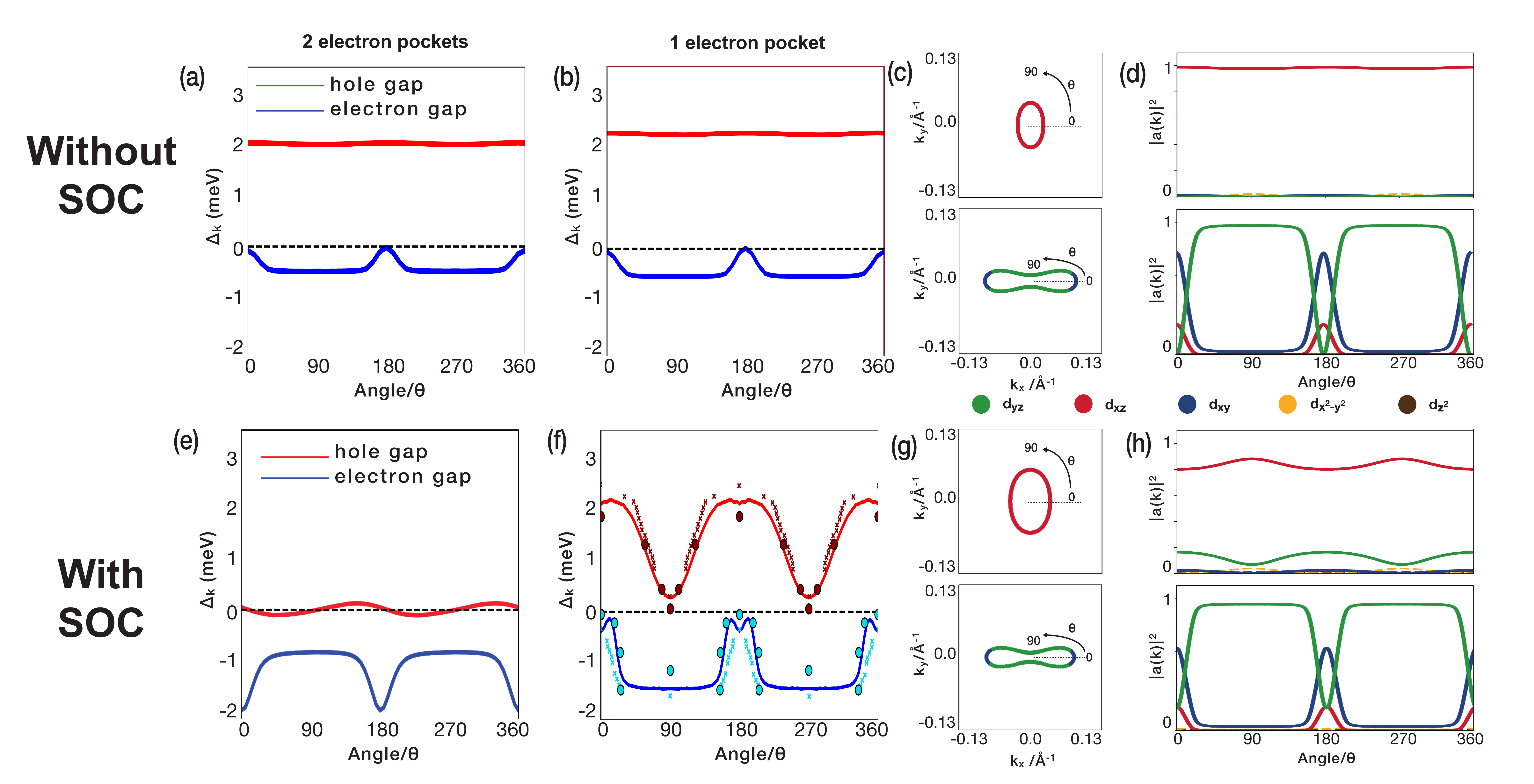}
 	\caption{\label{SOC_Comp} Comparison of the gap structure with and without spin orbit coupling. a) Momentum dependence of the gap structure without spin orbit coupling and including both electron pockets. b) Gap structure without spin orbit coupling and only including the $(\pi,0)$ electron pocket. c) Fermi surface maps of the hole (top) and electron pocket (bottom) without spin orbit coupling. d) Corresponding orbital characters as a function of angle around the pocket. e-h) The same as a-d) except with the inclusion of spin orbit coupling. }
 \end{figure}
 
 In Fig. \ref{SOC_Comp} We present the results of the linearized gap equation without the inclusion of spin orbit coupling, following the methodology of Ref. \cite{Graser2009}. We then compare this with the results from the linearized gap equation in the presence of spin orbit coupling as presented in the main text.
 
 Without spin orbit coupling, we obtain an isotropic gap dispersion at the hole pocket, which follows the $d_{xz}$ weight. Here, the momentum dependence of the gap at the $(\pi,0)$ electron pocket follows the $d_{yz}$ weight. There is very little change in the calculated gap structure by the removal of the $\mathbf{k}$-states associated with the $(0,\pi)$ electron pocket due to the fact that there is no possible intra-orbital scattering vectors between the hole and electron pocket. 
 
 By including spin orbit coupling, we mix the $d_{xz}$ and $d_{yz}$ states at the hole pocket, and induce a small portion of $d_{yz}$ orbital character at the hole pocket. This small addition has a large effect on the observed gap dispersion as it introduces intra-orbital scattering in the $d_{yz}$ channel between the hole and $(\pi,0)$ electron pocket. Therefore, not only is it important to accurately describe the band dispersions of FeSe, but it is also equally important to correctly describe the orbital character of these bands as well. This can only be achieved by the inclusion of spin orbit coupling and a nematic order parameter that well describe the shape of the experimental Fermi surface.

\end{document}